\documentclass[twocolumn]{aastex631}

\defcitealias{Buisson2018}{B18}

\shorttitle{Coronal temperature variation in ESO 103--035}
\shortauthors{Barua et al.}
\graphicspath{{./}{figures/}}

\begin{document}

\title{Evidence for coronal temperature variation in Seyfert 2 ESO 103--035 using {\textit {NuSTAR}} observations}

\correspondingauthor{V. Jithesh}
\email{vjithesh@iucaa.in}

\author[0000-0002-5248-2422]{Samuzal Barua}
\affiliation{Department of Physics, Gauhati University, Jalukbari, Guwahati-781014, Assam, India}

\author[0000-0002-6449-9643]{V. Jithesh}
\affiliation{Inter-University Centre for Astronomy and Astrophysics (IUCAA), PB No.4, Ganeshkhind, Pune-411007, India}
\affiliation{Department of Physics, University of Calicut, Malappuram-673635, Kerala, India}

\author[0000-0002-7609-2779]{Ranjeev Misra}
\affiliation{Inter-University Centre for Astronomy and Astrophysics (IUCAA), PB No.4, Ganeshkhind, Pune-411007, India}

\author[0000-0003-1589-2075]{Gulab C Dewangan}
\affiliation{Inter-University Centre for Astronomy and Astrophysics (IUCAA), PB No.4, Ganeshkhind, Pune-411007, India}

\author[0000-0003-0594-4521]{Rathin Sarma}
\affiliation{Department of Physics, Rabindranath Tagore University, Hojai-782435, Assam, India}

\author[0000-0001-6328-4512]{Amit Pathak}
\affiliation{Department of Physics, Banaras Hindu University, Varanasi-221005, India}

\author[ 0000-0002-3448-8150]{Biman J Medhi}
\affiliation{Department of Physics, Gauhati University, Jalukbari, Guwahati-781014, Assam, India}




\begin{abstract}

We report flux-resolved spectroscopic analysis of the active galactic nucleus (AGN) ESO 103--035 using  \textit{NuSTAR} observations. Following an earlier work, we fit the spectra using a thermal Comptonization model with a relativistic reflection component to obtain estimates of the coronal temperature for two flux levels. The coronal temperature was found to increase from 24.0$^{+6.8}_{-3.4}$ to 55.3$^{+54.6}_{-7.2}$ keV (errors at 1-$\sigma$ confidence level) as the flux increased from $9.8$ to $11.9 \times 10^{-11}$ erg cm$^{-2}$ s$^{-1}$ in the 3--78 keV band. A marginal variation in the high energy photon index allows for both, a non-varying optical depth and for the optical depth to have varied by a factor of $\sim$2. This is in contrast to a previous work on \textit{NuSTAR} flux resolved spectroscopy of the AGN, Ark 564, where the temperature was found to decrease with flux along with a $10$\% variation in the optical depth. The results maybe understood in a framework where AGN variability is either dominated by coronal heating variation leading to correlated increase of temperature with flux and the opposite effect being seen when the variability is dominated by changes in the seed photon flux. 
\end{abstract}



\section{Introduction}
\label{section1}
Strong and rapid variation in the X-ray emission is
one of the foremost characteristics of active galactic nuclei \citep[AGN;][]{McHardy1987, Lawrence1993, Green1993, Fabian1999, Nandra2001, Boller2003}. 
AGN host a super-massive black hole \citep[SMBH;][]{Peterson1997, Fabian1999, Beckmann2012}, 
which powers the activity through 
accretion of matter \citep{Salpeter1964, Lynden-Bell1969, Rees1984},
at a rate sometimes close to the Eddington limit \citep{Fabian1999}.

The hard X-ray emission from AGN is a consequence of inverse-Compton scattering of soft photons from
an accretion disk  around the SMBH by  hot thermal electrons in a corona located above the disk \citep{Sunyaev1979, Haardt1993, Merloni2003}. The X-ray emission sometimes strike the optically thick accretion disk and gets reflected, resulting in an iron K-line emission peaked at around 6--7 keV and a Compton reflection hump beyond $\sim$20 keV \citep[e.g.][]{George1991}. Furthermore, these reflection features get modified by the effects of strong gravity. 

The coronal temperature, in principle, can be determined by  fitting  broadband X-ray spectrum using Comptonization and reflection components. Nevertheless, until recently, it was challenging to conduct such  measurements due to the low sensitivity of the detectors at high energies and the complexities that arise while the fitting a relativistically blurred reflection component. 

\begin{table*}
	\centering
	\caption{Fitted parameters from the time-averaged {\it NuSTAR} FPMA and FPMB spectra}
	\label{tab:Table1_table}
	\setlength{\tabcolsep}{1.0pt}
	\begin{tabular}{lcccccccccc}
		\hline
		\hline
		N$_H$ &        $\Gamma$        &          $A_{Fe}$           &       $kT$$_{e}$        &        log$\xi$        &        $\theta$         &           R            & $R_{in}$ &           Norm           &           $\rm F_{3-78\,keV}$           & $\chi^2_{r}$ \\
		(10$^{22}$ cm$^{-2}$)                               &                        &        (solar)         &          (keV)          & [log(erg cm s$^{-1}$)] &        (degree)         &            &     ($r_{g}$)       &       (10$^{-4}$)        & ($\rm 10^{-11} erg\, cm^{-2}\, s^{-1}$) &    /d.o.f    \\ \hline
		16.2$_{-0.53}^{+0.59}$                              & 1.73$^{+0.02}_{-0.02}$ & 0.99$^{+0.14}_{-0.13}$ & 20.20$^{+2.88}_{-1.83}$ & 2.80$^{+0.07}_{-0.08}$ &                   $<17$ & 0.18$^{+0.01}_{-0.01}$ & 24.43$^{+6.02}_{-3.16}$ & 1.77$^{+0.09}_{-0.07}$  & 9.89$^{+0.03}_{-0.03}$ & 0.96/1291  \\\\

		15.96$_{-0.34}^{+0.56}$ & 1.79$^{+0.03}_{-0.02}$ & -- & 32.53$^{+7.63}_{-4.15}$ & 2.12$^{+0.08}_{-0.19}$ & -- & 0.29$^{+0.07}_{-0.05}$ & -- & 1.91$^{+0.06}_{-0.04}$ &  10.47$^{+0.03}_{-0.03}$ &  --   \\
		\hline
	\end{tabular}
\tablecomments{The spectra from both $\sim$27 ks and $\sim$42 ks observations are fitted jointly using {\tt relxillCp} model in 3--78 keV band. The model includes  absorption using \texttt{zTBABS}. From left to right the model parameters are: intrinsic absorption (N$_H$), photon index ($\Gamma$), iron abundance ($A_{Fe}$), temperature of coronal electrons ($kT$$_{e}$), log of the ionization parameter  ($\xi$), inclination of the accretion disk ($\theta$), reflection fraction (R), the inner disk radius ($R_{in}$), normalization, unabsorbed 3--78 keV flux ( $\rm F_{3-78\,keV}$) and reduced $\chi^2$ with degrees of freedom. The first row lists the free parameters from the  $\sim$27 ks data, while the second row lists the parameters for the $\sim$42 ks data. The inclination, iron abundance and inner disk radius are tied between the two observations.}
\end{table*}

However, these difficulties have been overcome by data from {\it Nuclear Spectroscopic Telescope Array} \citep[{\it NuSTAR};][]{Harrison2013}, which has  X-ray detectors operating in  3--79 keV energy range. Using focusing optics, {\it NuSTAR} provides an angular resolution (FWHM) of 18$\arcsec$, and has a spectral resolution (FWHM) of 0.4  and 0.9 keV at 10 and 60 keV, respectively. It is sensitive enough to make a 3-$\sigma$ detection in a 10$^6$ s exposure of a source with flux $\sim 2 (\sim 10) \times 10^{-15}$ erg cm$^{-2}$ s$^{-1}$ in the 6--10 (10--30) keV band \citep{Harrison2013}. As demonstrated by \citet{Garca2015}, these specifications are sufficient for determining the high-energy cut-off, which in turn allows one to measure the coronal temperature through spectral fitting. Indeed, {\it NuSTAR} observations of a number of AGN have provided the opportunity to determine their corona temperature. For a number of AGNs, the high energy cut-off has been estimated using {\it NuSTAR} data as listed in Table 1 of \citet{Fabian2015}. These high energy cut-offs can then be used to estimate the coronal temperature \citep[e.g.][]{Mid2019} which in turn allows for testing whether pair-production is important for these systems or not \citep{Fabian2015}. For a thermal plasma, there is a maximum threshold temperature beyond which a pair cascade would occur and this threshold depends on the compactness parameter $l = L\sigma_T/Rm_ec^3$, where $L$ is the luminosity and $R$ is the size of the corona \citep{Svensson1984, Zdziarski1985}. The estimated temperatures for some of the sources are close to the pair-production threshold value, which suggests that the temperature is perhaps being regulated by pair production. The inclusion of non-thermal electrons in the corona decreases the pair production threshold temperature, and thus the sources with cooler temperatures may also be regulated by the same mechanism \citep{Fab2017}. Correlation between the cut-off energy and spectral index over long time-scales have been reported \citep[e.g.][]{Zog2017, Tor2018}. Direct temperature measurements have been made, for example for 3C 382 \citep{Bal2014}, but these were typically at high values ($> 100$ keV). Recently, there have been reports of direct measurement of low coronal temperatures in AGN. The AGN Ark 564 was found to have a temperature of $\sim$ 15 keV \citep{Kara2017}. Temperature estimates have been reported for a few other AGN, ESO 103--035 ($kT_e \sim$ 22 keV), IGR 2124.7$+$5058 \citep[$kT_e \sim$ 20 keV;][hereafter \citetalias{Buisson2018}]{Buisson2018}, 2MASS~J1614346+470420 ($kT_e \sim$ 45 keV) and B1422+231 \citep[$kT_e \sim$ 28 keV;][]{Lanzuisi2019}.

\begin{figure*}
	\centering
	\includegraphics[width=17.7cm,height=6.5cm]{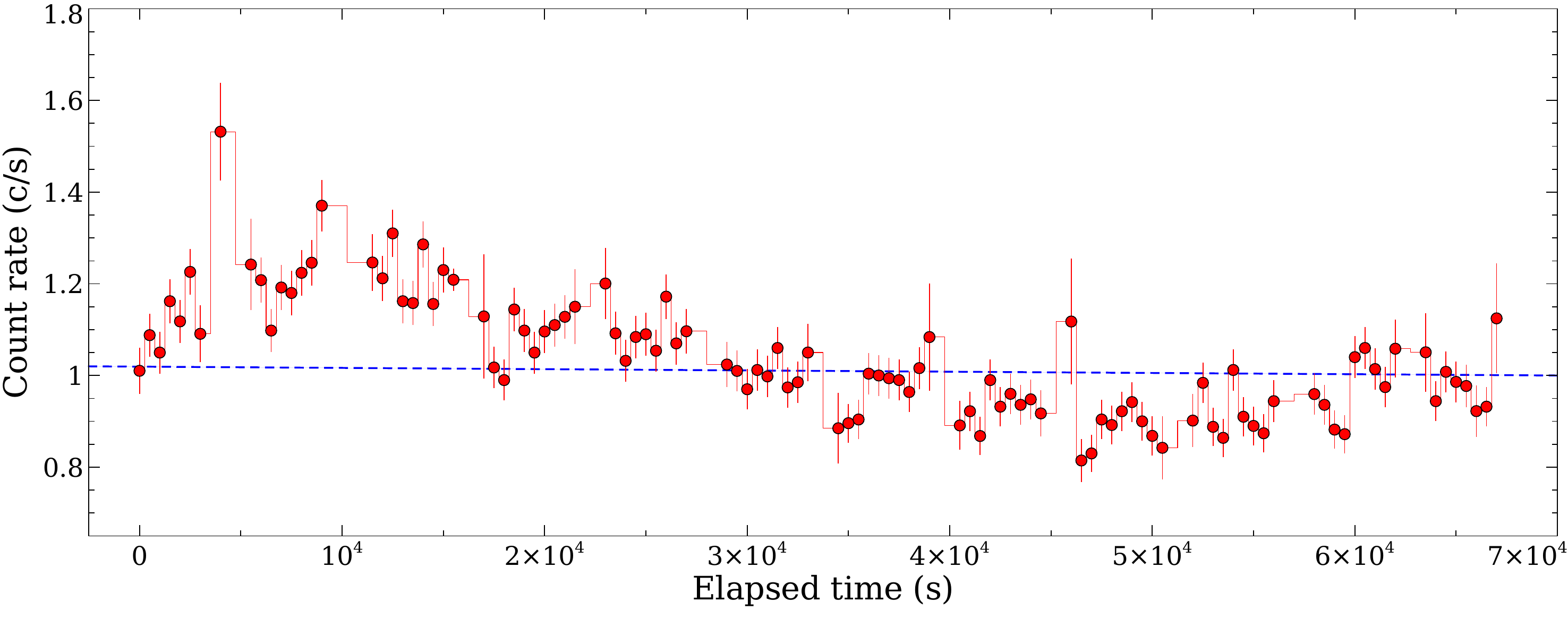}
	\caption{The $\sim$42 ks \textit{NuSTAR} light curve of ESO 103--035 extracted with a bin size of 500 s. The blue dotted horizontal line divides the light curve into two flux states, low: 0--1.02 c/s and high: 1.02--1.6 c/s flux states.}
	\label{fig:lcurve_figure}
\end{figure*}

AGN variability is observed in flux and continuum shape changes which are sometimes correlated. Analysing such variability in detail, provides  physical insight into the AGN. One of the  important variability behavior that has been observed in both AGN and X-ray binaries, is that the spectral index of the X-ray continuum correlates with the soft X-ray flux \citep{Haardt1997, Zdziarski2003, Sobolewska2009}. The X-ray continuum steepens as the source flux increases which could be a consequence of inverse-Comptonization process \citep{Zdziarski2003}.  Since radiative cooling is efficient, the corona should be continuously heated via some phenomenon so as to sustain it at a high temperature \citep{Fabian2015}. Now, if the input soft photon flux increases, the radiative cooling of the corona becomes more effective and  in case the heating rate is a constant, the coronal temperature should naturally decrease.
If the optical depth of the corona remains the same, such a decrease in the temperature would result in
hardening of the spectrum and hence this may explain the steepening of the spectrum as the flux
increases. Until recently, there were only  indirect ways  to infer that the coronal temperature
indeed varies with flux. For example, \citet{Wilkins2015} inferred the change in temperature
by showing that the corona has undergone an expansion.

In \citet{Samuzal2020}, we performed flux resolved spectroscopy of {\textit{NuSTAR} data of the AGN Ark 564,
and showed that the coronal temperature decreased with increasing flux. Furthermore, the variation
was accompanied by an increase in optical depth. ESO 103--035 is another suitable AGN that belongs to the class of Seyfert 2 type, whose {\textit{NuSTAR}} data  provides the opportunity to test the variation of coronal temperature with flux. Recently, \citetalias{Buisson2018} have analyzed 3--78 keV X-ray spectra of ESO 103--035 using two  {\textit{NuSTAR}} observations conducted in 2013 and 2017, with exposure times of 27.3 and 42.5 ksecs, respectively. The spectral analysis with the {\it NuSTAR} data provided direct observational evidence for a low-temperature corona, $\sim 22$ keV in ESO 103--035. 

ESO 103--035 \citep[z = 0.0133 $\pm$ 0.0003;][]{Phillips1979} was discovered as a powerful source of X-ray emission with HEAO-A2 \citep{Marshall1979, Piccinotti1982}. The X-ray spectrum of ESO 103--035 revealed a strong iron K$\alpha$ line \citep{Wilkes2001} along with some absorption features \citep{Phillips1979}. From the {\textit{BeppoSAX} observation in 1996, \citet{Akylas2001} reported a spectral index of $\sim$1.87, and an iron emission line peaked at 6.4 keV with a width of 0.3 $\pm$0.1 keV. In the {\textit{BeppoSAX} observation, the continuum spectrum could be  described by a power law with a high energy cut-off at 29 $\pm$10 keV \citep{Wilkes2001}. This observation also revealed that the source is highly absorbed with a column density N$_{H}$ = 1.79 $\pm$ 0.09 $\times$ $10^{23}$ cm$^{-2}$. The timing properties of the source (i.e. the shape of the power density spectrum) provided a black hole mass estimate of M$_{BH}$ = 10$^{7.1 \pm 0.6}$ M$_\odot$ \citep{Czerny2001}. 

In this work, we re-analyze the 27.3 ks and 42.5 ks \textit {NuSTAR} observations  to investigate the flux dependent spectral variability of Seyfert 2 ESO 103--035. In \S 2, we describe the observation used and data reduction. In \S 3, we present the results of the spectral fitting analysis and  discuss the results in \S 4.

\section{Observation and data reduction}

The \textit{NuSTAR} satellite observed the Seyfert 2 ESO 103--035 in 2013 and 2017 for exposure of 27.3 ks (Observation ID:60061288002) and 42.5 ks (Observation ID: 60301004002), respectively. 
The source has also been observed by {\it Swift}, simultaneously to both {\it NuSTAR} observations. However, the {\it Swift} exposure times are short, $\sim$ 6.8 ks and $\sim$ 2 ks for the $\sim$ 27 ks and $\sim$ 42 ks of {\it NuSTAR} data sets, respectively. These short exposure data are insufficient for flux resolved analysis and hence we have not used the {\it Swift} data. The \textit {NuSTAR} data processed with \textit {NuSTAR} data analysis pipeline ({\sc nupipeline}) in order to produce calibrated and filtered data products. For this standard pipeline processing, we used \textit {NuSTAR} data analysis software ({\sc nustardas}) v1.8.0 and CALDB version 20171002. From the cleaned event files, all required final products such as light curves, spectra were extracted using {\sc nuproducts} script. We extracted a circular source region with a 60 arcsec radius and a larger circular background region with a 90 arcsec radius that was away from from the source region (as used in \citetalias{Buisson2018}). SAA filtering was applied using ``saacalc=2 saamode=optimized tentacle=yes". The extracted regions for both the source as well as background were the same for two \textit{NuSTAR} instruments, Focal Plane Module A and Focal Plane Module B (FPMA and FPMB). The FPMA and FPMB spectra were grouped using the {\sc grppha} tool so as to have 50 counts in each spectral bin. 

\section{SPECTRAL FITTING}
\subsection{Time-averaged spectrum}

We first re-analysed the time averaged spectra of the two observations as has been done by \citetalias{Buisson2018}.
While, \citetalias{Buisson2018} fitted the spectra with different physical models, here we focus only on the
thermal Comptonization model with a relativistic reflection \texttt{relxillCp}\footnote{\url{http://www.sternwarte.uni-erlangen.de/~dauser/research/relxill/}}, since we are interested in estimating
the coronal temperature. This model  assumes that primary component  is due to inverse-Comptonization represented by the {\sc xspec} model \texttt{nthComp} \citep{Zdziarski1996, zycki1999}. Additionally, we use the absorption model \texttt{zTBABS} \citep{Wilms2000} to account for the intrinsic absorption in the source. We estimate the errors and the confidence contours using the  Markov chain Monte Carlo (MCMC) analysis for which we use the {\sc xspec\_emcee} code developed by Jeremy Sanders\footnote{\url{https://github.com/jeremysanders/xspec_emcee}} \citep{Sanders2013}, which is a Python implementation of Goodman \& Weare's Affine Invariant Markov chain Monte Carlo (MCMC) Ensemble sampler.  We use the default number of walkers 50 with 15000 iterations each, then burned the first 1000 in order to estimate errors from the steady chain. We quote 1-$\sigma$ errors on best-fit parameters corresponding to 68\% confidence level everywhere unless otherwise specified. To take into account possible difference in the effective area of the  FPMA and FPMB instruments, a constant multiplier was applied to the model for FPMB spectrum.

We first reproduce the results obtained in \citetalias{Buisson2018} by fitting the spectra of the two observations and found that the spectral parameters obtained were consistent with theirs. For the $\sim 42$  and $\sim 27$ ksec observation, we obtained the best fit values of the coronal temperature to be $kT_{e} = 25.9^{+12.4}_{-4.7}$ and $kT_{e} = 50.2^{+38.4}_{-13.5}$ keV, respectively as compared to $22_{-6}^{+19}$ and $> 20$ keV reported by \citetalias{Buisson2018}. We could also broadly constrain the inner disk radius $R_{in} \sim 25^{+6}_{-8} r_g$ while \citetalias{Buisson2018} report and upper limit of $17 r_g$.  The slight difference in the parameter ranges could be due to the use of MCMC technique and/or due to the presence of multiple close local minima in the $\chi^2$ space caused by the complexity of the model. Like other Seyfert 2 AGN, a narrow Iron K-$\alpha$ line emission has been reported for the ESO 103--035 \citep{2009ApJ...705..454N, 2010ApJ...725.2381L}. However, this emission line is weaker than what is typically observed in Seyfert 1 AGN \citep{2010ApJ...725.2381L}. These may be the reason why here in this work, we do not see any significant improvement in the spectral fitting upon adding a narrow reflection component.

Next, we fitted the spectra from the two observations jointly, keeping the Iron abundance ($A_{Fe}$), the disc inclination angle ($\theta$) and the inner radius ($R_{in}$) to have the same values, while allowing for the rest of the free parameters to vary independently. This is physically reasonable since these parameters are not expected to vary. The best-fit free spectral parameters from the MCMC analysis are listed in Table~\ref{tab:Table1_table}. The spectral fitting is rather insensitive to the other parameters of the model which describes the reflection component \citepalias{Buisson2018} and hence  were fixed at nominal values i.e. the outer radius to $400 r_g$, the emissivity index to 3 and the spin of the black hole to $0.99$.

 \begin{figure}
 	\hspace{-4.1mm}
 	\includegraphics[width=8.7cm, height=7.2cm, angle=0]{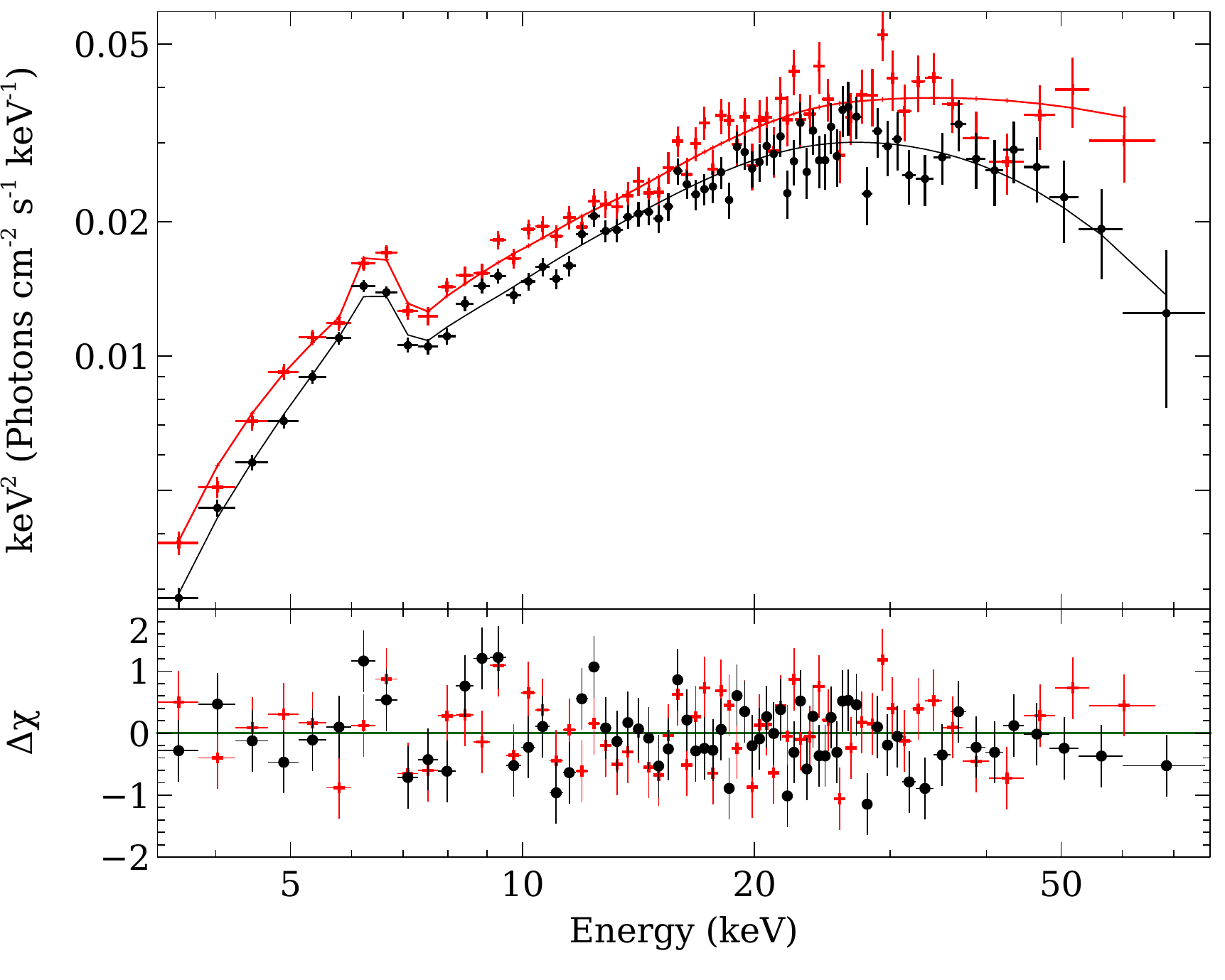}
 	\caption{Unfolded \textit{NuSTAR} FPMA spectra of ESO 103--035 from two flux states. The black circles and red plus represent the low and high flux states, respectively. These spectra are fitted with \texttt{relxillCp} model.}
 	\label{fig:spectra_figure}
 \end{figure}

This joint fitting allows for a better constrain on the electron temperature for both observations. The estimated temperatures are constrained to be at $20.2^{+2.9}_{-1.8}$ and $32.5^{+7.6}_{-4.2}$ keV, for the 27 ks and 42 ks observations, respectively. In the joint fit, some other parameters like the Iron abundance are also better constrained, however the best fit values are roughly similar to the ones obtained by the individual fits and by \citetalias{Buisson2018}.



 \begin{figure*}
 	\centering
 	\includegraphics[width=\columnwidth]{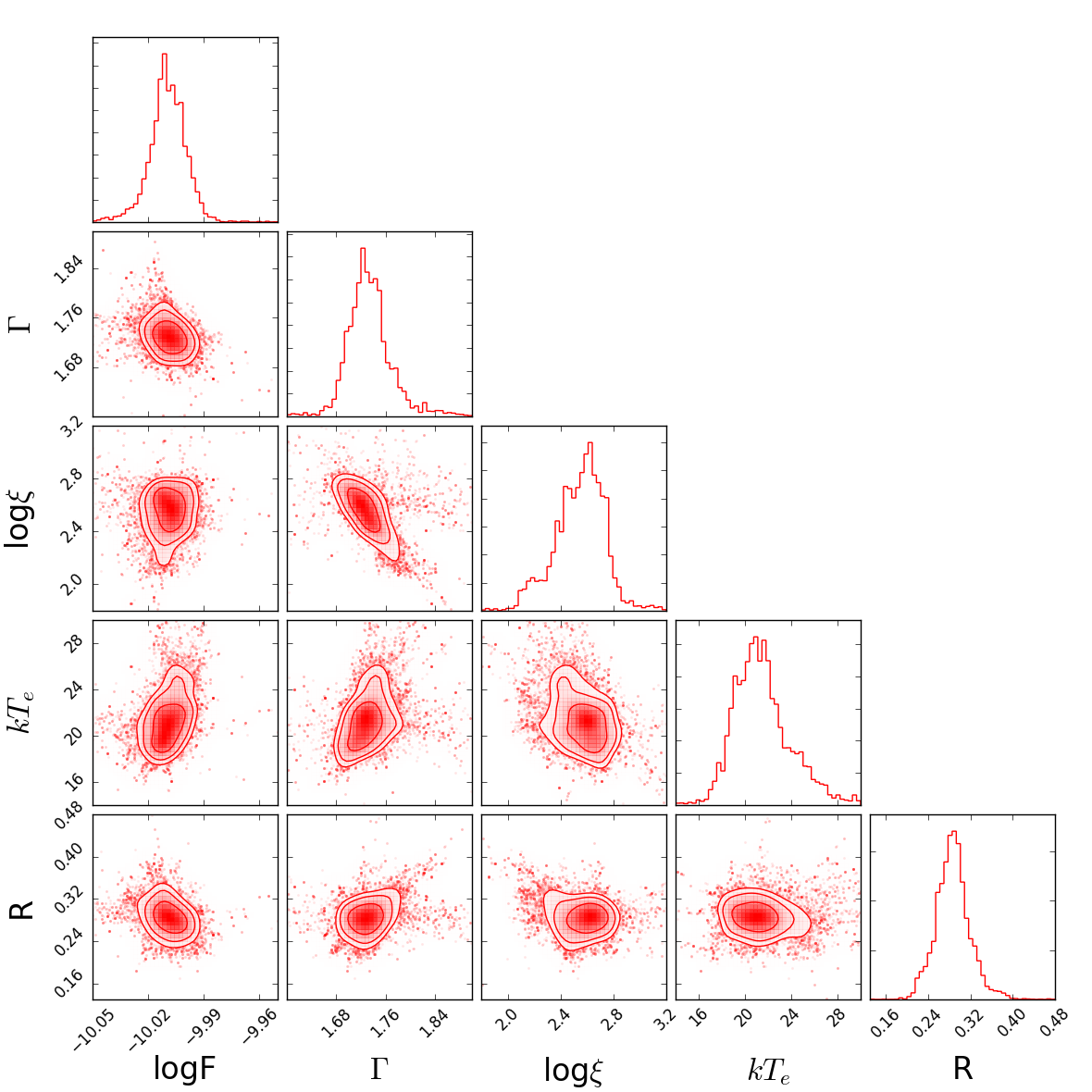}
 	\hspace{0.15cm}
 	\includegraphics[width=\columnwidth]{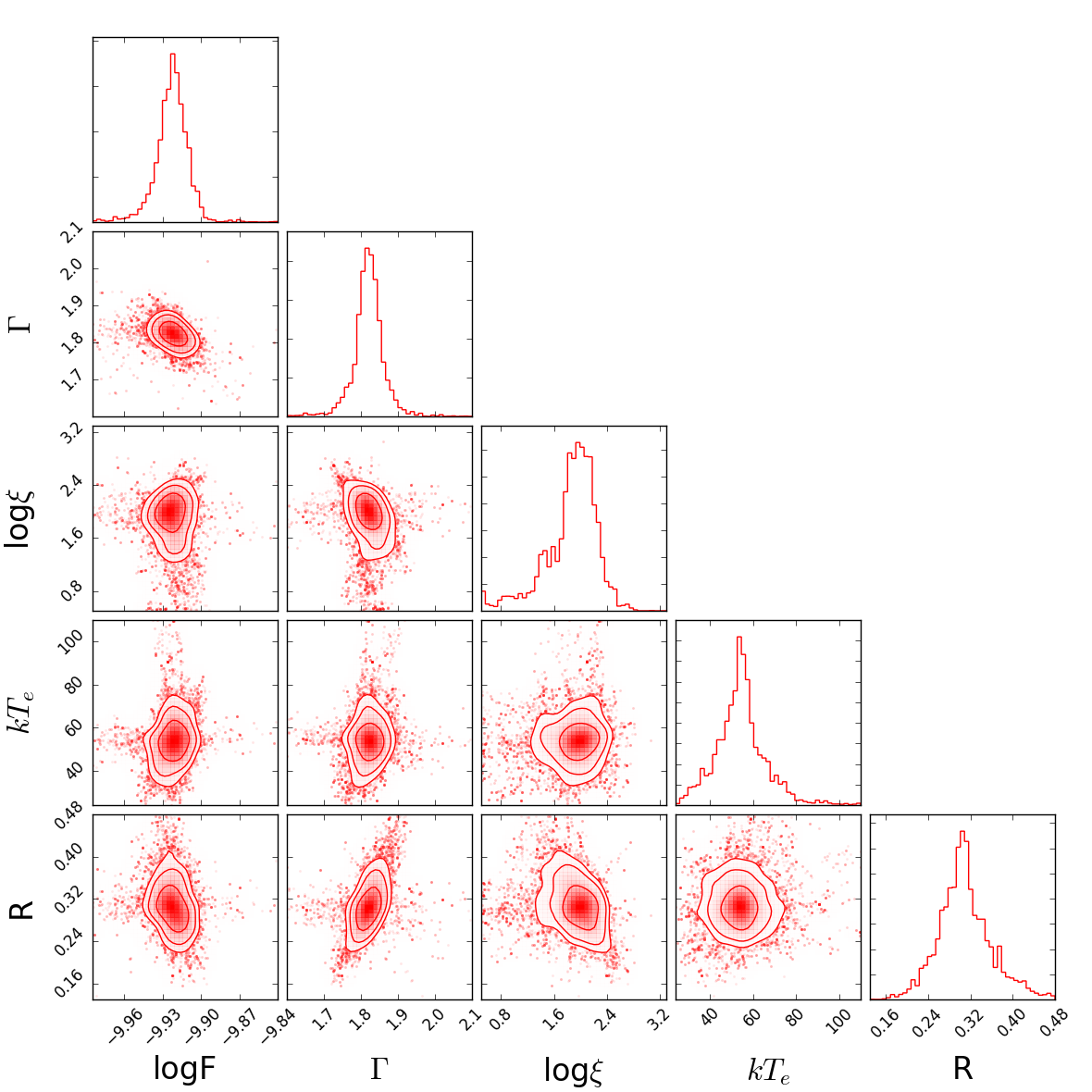}
 	\caption{Corner plots of spectral parameters obtained from the analysis of flux-resolved spectra. The left and right hand panels represent MCMC results for low and high flux states, respectively.}
 	\label{fig:corner_fl}
 \end{figure*}	

\begin{figure}
	\includegraphics[width=\columnwidth]{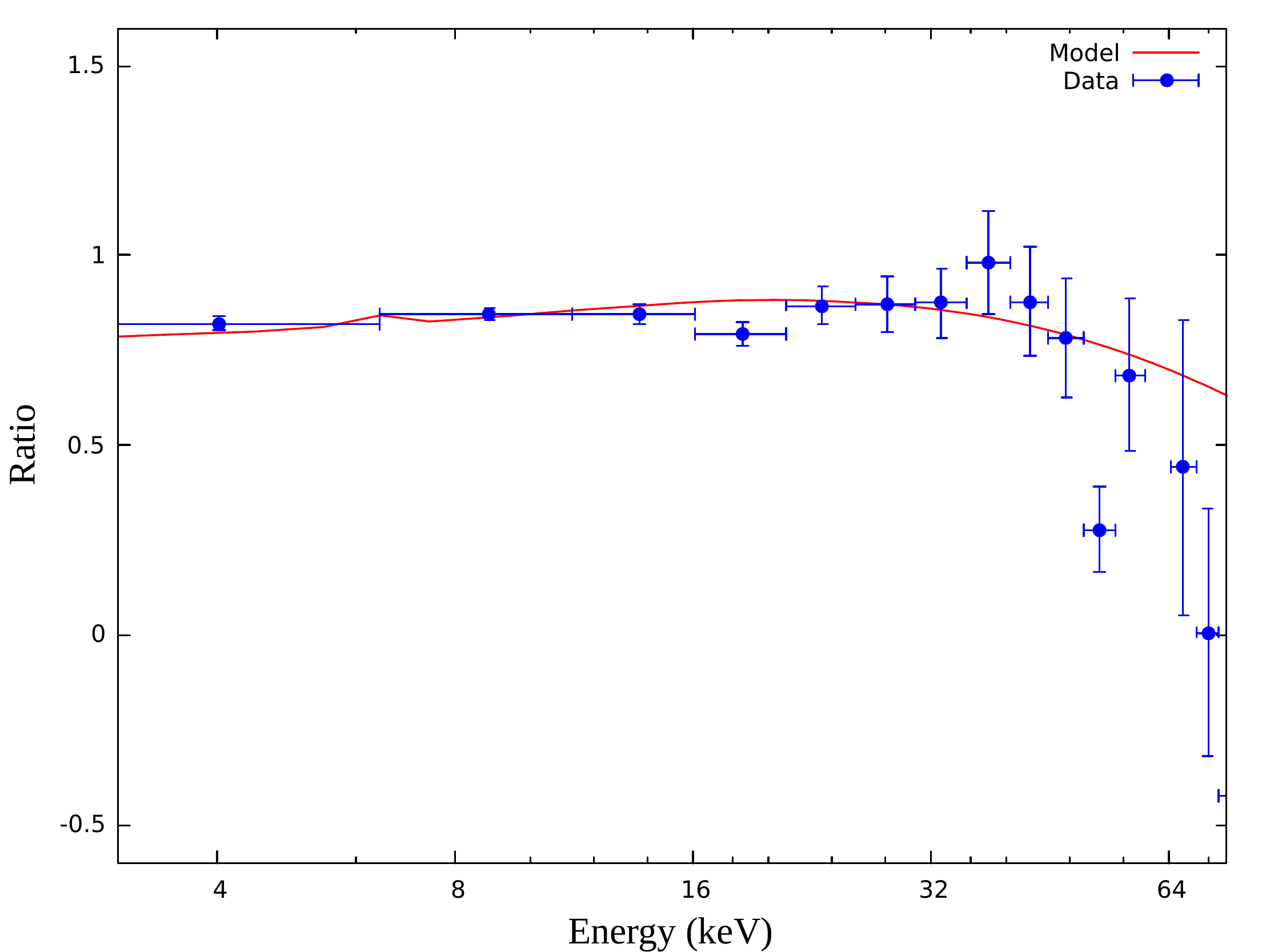}
	\caption{The ratio of the \textit{NuSTAR} data from the low and high flux states. The blue circles represent the ratio of the spectra from high to low flux states, whereas the red line represents the model ratio of the same.}
	\label{fig:ratio_figure}
\end{figure}  


\subsection{Flux-resolved spectra}
\label{section3.2}
To study the flux dependent coronal temperature variation in ESO 103--035,
we obtained  flux-resolved spectra by splitting the  42.5 ks \textit {NuSTAR} observation into two different flux states, which are  denoted here as low and high flux states. These low and high flux states consisted of data when the count rate was in the range  0--1.02 c/s and 1.02--1.6 c/s, respectively as shown in Figure \ref{fig:lcurve_figure}. This resulted in the high (low) flux state to have an exposure of $\sim 16$ ksec ($\sim 27$ ksec) and total counts of $\sim 13300$ ($\sim 18500$). The $\sim$27 ks observation of ESO 103-035 does not exhibit significant flux variation in the light curve (see Figure 1 of \citetalias{Buisson2018}), and hence was not considered for flux-resolved spectroscopy.

The average count rates for  low and high flux states are 0.674 $\pm$ 0.005 and 0.806 $\pm$ 0.007 c/s, respectively. Similar to the spectral modeling of the time-averaged spectrum as discussed in the previous section, the 3--78 keV flux-resolved spectra were fitted using the same relativistic reflection model \texttt{relxillCp}. The iron abundance ($A_{Fe}$), Inner disk radius ($R_{in}$) and the absorption column density were fixed to the values  obtained from the time-averaged spectrum. Also, we fixed the inclination angle ($\theta$) to be 10 degree as it was constrained to be less than 17 degree in the time-averaged analysis. The unfolded spectra and residuals are shown in Figure \ref{fig:spectra_figure}. The unabsorbed flux in the 3--78 keV range was estimated using the {\sc xspec} model \texttt{cflux} and  the free spectral parameters are listed in Table~\ref{tab:Table2_table}. The spectral analysis of these spectra indicates the variation of the coronal electron temperature, which increased from $24.0^{+6.8}_{-3.4}$ to $55.3^{+54.6}_{-7.2}$ keV as the flux increased. This temperature variation is seen to accompanied by a marginal increase of the photon index, from $1.76^{+0.02}_{-0.03}$ to $1.81^{+0.06}_{-0.02}$. In order to examine the degeneracy between the spectral parameters we constructed corner plots from the MCMC analysis and show them in  Figure \ref{fig:corner_fl}. It is seen that though some pairs of parameters show moderate to strong degeneracy still all of them are well constrained. To show the spectral variation in a model independent way, the ratio of the high to low spectra is shown in  Figure \ref{fig:ratio_figure} which indicates a  curvature at high energies, implying an increase of the coronal temperature as the flux increases. We further quantify the significance of the coronal temperature variation result by fitting a constant to the two temperature values which gives a $\chi^{2}$ of 5.25 for one degree of freedom. This implies a null hypothesis probability that the temperatures are same to be less than 0.02 or in other words the result that the temperatures are different is at 98 \% confidence level.

We also fitted the high and low flux state spectra joint by keeping the temperature tied, which resulted in a $\chi^2$/dof = 871.9/923 as compared to when they were kept free, $\chi^2$/dof = 866.5/921. An F-test gives the significance of the temperature variation at $\sim 95$\%. Instead of {\tt relxillCp}, if we use {\tt relxill} (which incorporates a cut-off power-law continuum instead of a thermal Comptonization one), the cut-off energy was found to be larger for the high flux state, $69.5^{+16.6}_{-11.4}$ keV as compared to the low flux state, $39.3^{+4.9}_{-2.9}$ keV, providing a supporting evidence to the observed temperature variation.
\begin{table*}
	\centering
	\caption{Fitted Parameters from the flux-resolved spectra of ESO 103--035 from $\sim$42 ks observation}
	\label{tab:Table2_table}
	\setlength{\tabcolsep}{6.6pt}
	\begin{tabular}{lccccccccccc} 
		\hline
		\hline
		Flux state& $\Gamma$& &  $kT$$_{e}$  && log$\xi$ &  R && Norm & $\rm F_{3-78\,keV}$ & $\chi^2_{r}$\\
		& && (keV)    &&[log(erg cm s$^{-1}$)] &   && (10$^{-4}$) & ($\rm 10^{-11} erg\, cm^{-2}\, s^{-1}$) & /d.o.f   \\
		\hline
		\vspace{0.4cm} 
		1 (Low)& 1.76$^{+0.02}_{-0.03}$ &&24.04$^{+6.80}_{-3.42}$ && 2.31$^{+0.09}_{-0.06}$ & 0.31$^{+0.06}_{-0.02}$ && 1.66$^{+0.10}_{-0.07}$   & 9.76$^{+0.04}_{-0.04}$ & 0.95/527 \\
		\vspace{0.2cm}
		2 (High)&1.81$^{+0.06}_{-0.02}$ &&55.3$^{+54.6}_{-7.2}$ && 2.17$^{+0.15}_{-0.33}$ & 0.28$^{+0.08}_{-0.04}$ && 2.29$^{+0.18}_{-0.05}$   & 11.87$^{+0.04}_{-0.04}$ & 0.93/395 \\
		\hline
	\end{tabular}
\tablecomments{The spectral fitting is done similar to the fits of time-averaged spectrum. The iron abundance ($A_{Fe}$), inner disk radius ($R_{in}$) and column density ($N_H$) parameters are fixed to the values obtained from the analysis of the time-averaged spectrum, while the inclination angle ($\theta$) is fixed at 10 degree.}
\end{table*}  
If instead of defining the high and low flux states as having count rate (CR) greater than or less than $\sim 1$ c/s, we choose $\sim 0.98$ c/s as the threshold, the best fit temperature of the high state turns out to be $46.2_{-9.0}^{+22.9}$ keV, while it is $21.3_{-1.6}^{+10.2}$ keV for the corresponding low state.  For a threshold of $\sim 1.1$ c/s, the high (low) state temperature turns out to be $103.4_{-11.2}^{+119.8}$ keV ($23.3_{-3.0}^{+6.7}$). Thus, the result that the coronal temperature increases with flux is not very sensitive to the demarking flux level.
If instead of two, we consider three flux states, defined as low 
(CR $< 0.98$ c/s), medium ($0.98 <$ CR $< 1.1$ c/s) and high (CR $> 1.1$ c/s), 
the best fit temperatures turn out to be $21.3_{-1.6}^{+10.2}$, 
$34.0_{-7.2}^{+18.0}$ and $103.4_{-11.2}^{+119.8}$ keV, respectively. The best fit 
temperatures of the two lower flux states are consistent with each other, 
and hence considering three flux states does not improve the statistical 
significance of the result. The errors quoted in this work are from the MCMC technique, which may 
sometimes be different from those obtained using the standard $\chi^2$ 
variance method. Using the standard method, the temperature estimates 
for the high and low flux states turn out to be $55^{+90}_{-16}$ and 
$24^{+7}_{-5}$ keV respectively, which although different from the 
values quoted in Table 2, are consistent with the result, that the 
temperature increased with flux.

\section{Discussion and Conclusion}
We have investigated the variability of the coronal temperature of ESO 103--035 using the \textit{NuSTAR} observations. Our study revealed that probably (at $>95\%$ confidence level) the coronal temperature increased by a factor of $\sim 2$ when the flux increased by $\sim 22$\%. While other spectral did not show any significant variation, the photon index ($\Gamma$) increased by roughly 0.1. Using the equation,

\begin{eqnarray}
\tau = \sqrt{\frac{9}{4} + \frac{3}{\theta \left[\left(\Gamma+\frac{1}{2}\right)^{2} - \frac{9}{4}\right]}} - \frac{3}{2}\ \ \ \ \
\end{eqnarray}
where $\theta=kT_e/m_ec^2$, we estimate the optical depth ($\tau$) of the corona to be $3.45^{+1.34}_{-0.76}$ and $1.85^{+2.98}_{-0.57}$ in low and high flux states, respectively. Thus, while the optical depth is not well constrained a large variation by a factor of two is also consistent. We note that the $\sim 27$ ksec observation with a flux of $\sim 9.9 \times 10^{-11}$ erg cm$^{-2}$ s$^{-1}$ and coronal temperature estimate of $20.2^{+2.9}_{-1.8}$ keV is consistent with the flux and temperature values obtained for the low flux resolved spectrum.

The results may be contrasted with that found for Ark 564 \citep{Samuzal2020}, where the coronal temperature was found to {\it decrease} with increasing flux accompanied by about 10\% increase in the optical depth. The different behavior of the two systems can be reconciled by considering that the flux variation in ESO 103--035 is due to changes in the heating rate of the corona, leading to a correlated variation of the coronal temperature and flux. On the other hand, for Ark 564, the flux variation could be driven by variation in the input seed photon flux, leading to the expected anti-correlation between temperature and flux. Both variations may well be accompanied by changes in the optical depth of the corona. 

It is interesting to note that there is a qualitative difference between the X-ray spectral shapes
of Ark 564 and ESO 103--035. While for Ark 564, the high energy photon index is $\sim 2.3$ that is
greater than 2.0, for ESO 103--035 it is $\sim 1.7 < 2.0$. This means that the flux of ESO 103--035 is
dominated by emission at high energies $\sim 30$ keV, while for ArK 564 it is dominated by low
energy emission. Thus, flux of ESO 103--035 would be more sensitive to changes in the coronal heating
rate than seed photon rate and the vice-versa would hold for Ark 564. This maybe a clue to the reason
why the two sources show different correlation of corona temperature with flux.

The results presented here along with the earlier result for Ark 564, suggest that AGN variability could
be dominated by either seed photon changes or coronal heating variation, with each process exhibiting a
different correlation of the corona temperature with flux. Clearly, the analysis has to be undertaken for
a larger number of sources, to draw conclusions regarding which process dominates and whether that depends
on the spectral properties of the source. Moreover, even for the same source, the two process could
dominate at different times. It is also likely that both coronal heating variation and seed photon
flux changes are active at the same time and perhaps can occur with a time difference. Here, one can
perhaps draw an analogy with black hole X-ray binaries  which also have a hot corona Comptonizing
low energy photons, to produce high energy X-ray emission. Detailed variability studies of X-ray binaries
have  revealed that both the seed photon flux and the coronal heating rate vary and they
do so after a time lag which depends on the time-scale of the variation \citep{Maq19, Jit19, Mud20, Jit21}. Flux resolved spectroscopy using sensitive instruments like {\it NuSTAR} for a larger sample of AGN with more continuous monitoring, will reveal the nature of the variability of these sources.

\begin{acknowledgments}
We thank the anonymous referee for the constructive comments and suggestions that improved this manuscript. SB acknowledges the IUCAA Visiting program. SB, RS \& RM acknowledge the SERB research grant EMR/2016/005835. This research has made use of data obtained from the High Energy Astrophysics Science Archive Research Center (HEASARC), provided by NASA's Goddard Space Flight Center, and the {\it NuSTAR} Data Analysis Software (NUSTARDAS) jointly developed by the ASI Science Data Center (ASDC, Italy) and the California Institute of Technology (Caltech, USA).  
\end{acknowledgments}

%

\vspace{5mm}
\facilities{NuSTAR}


\software{XSPEC\_EMCEE \citep{Sanders2013},  
          {NUSTARDAS},
          }






\end{document}